AHFE
International

# Sensor-based Data Acquisition via Ubiquitous Device to Detect Muscle Strength Training Activities

**Elizabeth Wianto[1], Hapnes Toba[2,\*], Maya Malinda[3] and Chien-Hsu Chen[4]**

[1]Bachelor Program in Visual Communication Design, Universitas Kristen Maranatha, Indonesia
[2,\*]Master Program in Computer Science, Universitas Kristen Maranatha, Indonesia
[3]Bachelor Program in Management, Universitas Kristen Maranatha, Indonesia
[4]Industrial Design Department, National Cheng Kung University, Taiwan

**ABSTRACT**

Maintaining a high quality of life through physical activities (PA) to prevent health decline is crucial. However, the relationship between individuals' health status, PA preferences, and motion factors is complex. PA discussions consistently show a positive correlation with healthy aging experiences, but no explicit relation to specific types of musculoskeletal exercises. Taking advantage of the increasingly widespread existence of smartphones, especially in Indonesia, this research utilizes embedded sensors for Human Activity Recognition (HAR). Based on 25 participants' data, performing nine types of selected motion, this study has successfully identified important sensor attributes that play important roles in the right and left hands for muscle strength motions as the basis for developing machine learning models with the LSTM algorithm.

## INTRODUCTION

The wide variation of people's health statuses combined with their preferences for PA is a complex issue. Its complexity then increased with age and the level of self-motivation (Schutzer and Graves, 2004). Discussing PA are consistently positively related to seven domain of positive aging experiences, which consists of: contains daily functioning, physical fitness, long-term physical health problems, heart health, weight, sleep, and subjective perceptions of health (Bone et al., 2023).

However, according to the National Institute of Aging, most people only focus on one type of activity and think they are doing enough (*Four Types of Exercise Can Improve Your Health and Physical Ability*, 2021) correlated with walking as the preferred activity. Therefore, it is advisable to do other types of exercises categorized as endurance, strength, balance, and flexibility. The selected types of PA in this paper focus on strength training, as this exercise is also recommended by the global health organization for the general population, conferring direct benefits to the musculoskeletal system in common disorders and healthy people (Maestroni et al., 2020).

This study aims to take advantage of the abundance of electronic health systems, such as smartphones and smartwatches, which are now ubiquitous across the whole population due to their components' affordability, thus bringing a new era of next-generation intelligent monitoring systems (Berenguer et al., 2016), (Birenboim and Shoval, 2018). Wearable electronic devices are conjectured as one of the triggers of habit-forming in society (Oulasvirta et al., 2012). Hence, together with the capabilities of the latest technology, people are now more conscious of their health and well-being (Meegahapola and Gatica-Perez, 2020). Embedded smartphone sensors can automatically detect the user's context (Mylonas et al., 2013). Therefore, in recent years, HAR and human behavior monitoring have gained much attention due to various feasible application domains, including techniques for the ambient environment or directly detecting human kinetic performance (Ramanujam, Perumal and Padmavathi, 2021), (Ronao and Cho, 2016), (Tsapeli and Musolesi, 2015), and was already been successfully implemented in outdoor activities and modified based on their target are known as a fitness tracker (Burton et al., 2018), (Cooper et al., 2018), (Mopas and Huybregts, 2020), (Steinert et al., 2018), (Vooris, Blaszka and Purrington, 2019).

To the extent of our knowledge, research on generic sensors to synchronize specific strength training moves is limited. In addition, reflecting on the global pandemic situation in the last three years, it has been seen that society also requires several indoor activities to be carried out statically in a limited space. Therefore, the main contribution of our study aims to initiate and detect specific motions known in strength training activities using motion sensors in smartwatches.

The data acquisition is limited to upper limb muscle strength motions. They are mainly performed using weight-bearing devices, such as dumbbells. We argue that ubiquitous electronic health systems, such as smartphones and smartwatches, are still affordable for people in developing countries such as Indonesia to support their health-conscious behavior. In this research, we also perform feature selection to highlight the importance of specific sensors to learn the motion prediction model.

## METHOD

The research was carried out with the following workflow: (1) Development of software for motion data collection; (2) Controlled data collection; (3) Extraction of data and feature selection for creating a prediction model; and. (4) Development of a prediction model using Long Short-Term Memory (LSTM) machine learning.

## Application Development

The software development process is carried out using an Agile approach that involves direct interaction with potential users. The main functional requirements of the software are as follows: (1) Users can enter their names when initiating motion capture; (2) Each motion repetition lasts approximately 7-8 seconds. To ensure the capture of all repetitions, 7 lines of sensor responses are stored for each second; (3) Users can select the motion name, click "start," and click "stop" when finished; (4) There is a dedicated button to send motion data to the central server; (5) Motion data is stored in a centralized database; and (6) Once all data is collected on the server, it can be exported in other formats for model generation through machine learning.

## Data Collection

The data collection process was conducted under controlled conditions with 25 students from the Faculty of Fine Arts and Design. To ensure uniformity in the motions performed by all participants, guidance and video observation were utilized from a YouTube channel (*Upper Body Exercises for Seniors and the Elderly, Strength training for seniors,* 2019). Nine specific motions were gathered from the video. Although the instructions in these videos are intended for seniors, data collection was carried out by younger individuals to test the application. This will also serve to educate the importance of maintaining muscle strength from an early age.

## Data Extraction and Feature Selection

The collected data is synchronized into the server and stored in a centralized database, as presented in Table 1.

**Table 1.** The dataset features

| Column Name | Data Type | Description | Technical Description |
|---|---|---|---|
| Respondent name | text | respondent's identification | To distinguish individuals who perform the motions in anonymous code. |
| Timestamp | datenum | timestamp of the data acquisition | Unix format timestamp, *i.e.,* representing the number of seconds passed since Jan 1st, 1970, UTC at midnight |
| Accelerometer | numeric | axis x, y, and z | A sensor that measures acceleration forces, detect motion orientation, and vibration in 3D. |
| Magnetometer | numeric | axis x, y, and z | A sensor that measures the strength and direction of magnetic fields. |
| Gyroscope | numeric | axis x, y, and z | A sensor that measures and detects rotational motion or angular velocity |
| Linear accelerometer | numeric | axis x, y, and z | A subset of accelerometers specifically measure linear acceleration along a single axis/direction. |
| Gravity | numeric | axis x, y, and z | Provide the device's orientation relative to the Earth's gravity vector and is used for applications such as screen rotation & virtual reality systems. |
| Euler | numeric | axis x, y, and z | A sensor that provides information about the device's orientation using Euler angles. |
| Quaternion | numeric | axis x, y, z, and w | A sensor that provides orientation information using quaternion representation. |
| Inverse quaternion | numeric | axis x, y, z, and w | A mathematical operation that produces the reciprocal of the original quaternion |

| Column Name | Data Type | Description | Technical Description |
|---|---|---|---|
| Relative orientation | numeric | axis x, y, and z | Identify device's physical orientation disregard of the Earth's reference coordinate system |
| Motion type | text | nine types of motion | Overhead press, bicep curls, lateral raise, overhead triceps, diagonal shoulder raise, forward punches, reverse fly, seated rows, and modified skull crushers, |
| Side | numeric | right & left-hand | Options: right and left |

After collecting data from the respondents on the server, the next step is to perform an analysis of important features. For feature analysis, a feature selection process using the filtering method is employed based on Pearson correlation calculations (Liu, 2019). The Pearson analysis helps identify which columns have strong correlations. We utilize the accelerometer attribute on the x-axis as the initial determinant feature for initiating a motion. This choice is based on the inherent characteristics of the sensor, which operates at a low level and is embedded in all modern mobile devices.

## Development of Prediction model

The motion prediction model is developed based on the Long Short-Term Memory (LSTM) algorithm. LSTM is a popular deep-learning algorithm suitable for making predictions and classifications related to time sequences. The LSTM algorithm's structure consists of a neural network and several different memory blocks known as cells. The hidden states of these cells are passed on to the next cell, creating a specific sequence range. This feature is particularly useful for predicting time series data as an indicator to predict specific classes (Siami-Namini, Tavakoli and Namin, 2019).

## RESULT

## Mobile Application and Smartwatch

The application is developed using the Flutter programming language in Android Studio. The application can be used on Android devices with a minimum OS version of 5.0 (API level 21). The resulting application can be installed on both smartphones and smartwatches. Through this application, the process of data acquisition can be performed anywhere as seen in Table 2.

## Data Collection

The data collection process was conducted in five sessions. Each session involves five students. Each student performed nine motions according to the instructions provided in the video under the authors' supervision. From these data collection sessions, a total of 69,088 rows of data were gathered. From the collected data, several observations are presented in Table 3. The average row number of each student to perform each motion will be used as the sequential period during the LSTM training. The mean of the left-hand and right-hand averages is then computed to determine the final timespan. In this case, the mean is (164+144)/2 = 154. Note that we rounded the number to 150 during the LSTM training. One of the reasons is compensation for the unused rows during the transition between motions.

**Table 2.** Screenshot of user interaction in the application

| Smartphone | Smartwatch | Description |
|---|---|---|
| 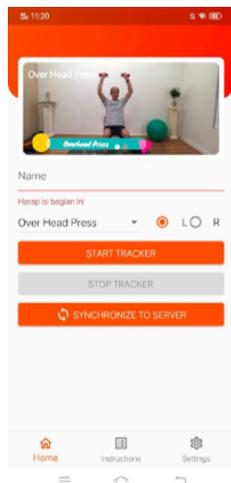 | 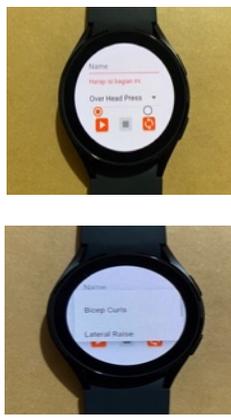 | On the home screen, users can input their names. After inputting the name field, users can choose:<br>1. the type of motion<br>2. whether they use the gadget in the right or left hand.<br>Users can start the data acquisition process by pressing the start tracker button. After finishing the recorded motion, participants press the synchronization button, to synchronize the data to the server. |

**Table 3.** Statistics of the collected dataset

| Characteristic | Left | Rounded | Right | Rounded |
|---|---|---|---|---|
| Number of total rows | 36,792 | 36,792 | 32,296 | 32,296 |
| Average row per student (25 students) | 1,471.68 | 1,472 | 1,291.84 | 1,292 |
| Average rows per student per motion (9 motions) | 163.52 | 164 | 143.54 | 144 |
| Average second per student per motion (per second ~ 7 rows) | 23.36 | 23 | 20.51 | 21 |
| Avg. second per repetition (assume 8 repetitions per motion) | 2.92 | 3 | 2.56 | 3 |

## Feature Selection

Based on the basic accelerometer attributes on the x-axis, feature selection was performed using Pearson correlation. The feature selection process was conducted separately for the left hand and right hand to ensure the union of attributes that have an impact on both hands. Table 4 provides the results of the feature selection process. There are 10 influential attributes for the left hand and 7 attributes for the right hand. As additional information, all the students who participated in the experiment were right-handed. An example of the data collection situation can also be seen in Figure 1.

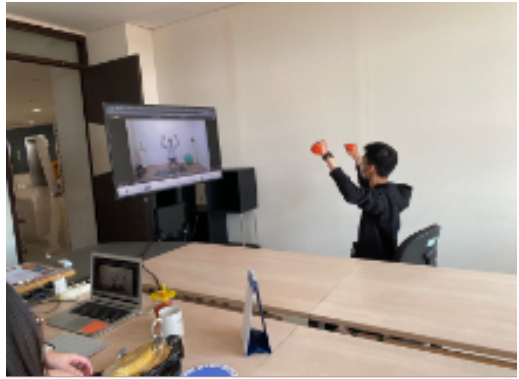

**Figure 1:** An example during the data collection process (source: authors' experiment results)

**Table 4.** The influential attributes of the left and right-hand

| Left-hand | Right-hand | Union |
|---|---|---|
| 1. accelerometer_x | 1.accelerometer_x | 1. accelerometer_x |
| 2. *linear_accelerometer_x* | 2.*magnetometer_x* | 2. *linear_accelerometer_x* |
| 3. gravity_x | 3.gravity_x | 3. gravity_x |
| 4. *euler_x* | 4.euler_z | 4. *euler_x* |
| 5. euler_z | 5.quaternion_z | 5. euler_z |
| 6. *quaternion_x* | 6.inverse_quaternion_z | 6. *quaternion_x* |
| 7. quaternion_z | 7.relative_orientation_z | 7. quaternion_z |
| 8. *inverse_quaternion_x* | | 8. *inverse_quaternion_x* |
| 9. inverse_quaternion_z | | 9. inverse_quaternion_z |
| 10.relative_orientation_z | | 10.relative_orientation_z |
| | | 11.*magnetometer_x* |

## LSTM Training

The LSTM model expects fixed-length sequences as training data. Each generated sequence contains 150 training examples as described in Table 3, and 11 features as given in Table 4. One-hot-encoding is used to determine the target motion classes during the training. A hold-out scenario is used with a proportion of 80% training and 20% testing data in 50 epochs with a ReLU activation function. Finally, the complete model contains 2 fully connected LSTM layers (stacked on each other) with 64 units each. The model architecture can be seen in Figure 2. The results of the training are encouraging. Each model of the right and left-hand dataset has an accuracy of 0.979 and 0.981 respectively. The training curves can be seen in Figure 3.

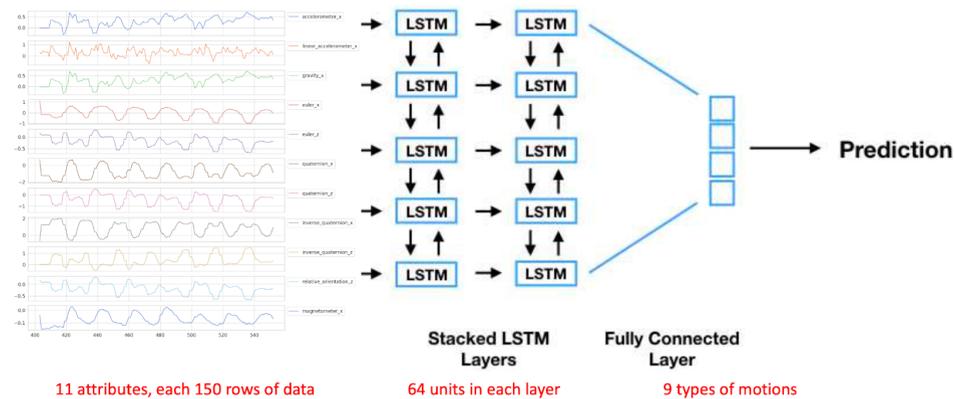

**Figure 2.** The generic LSTM architecture during training (source: adapted from (Zhu *et al.*, 2016))

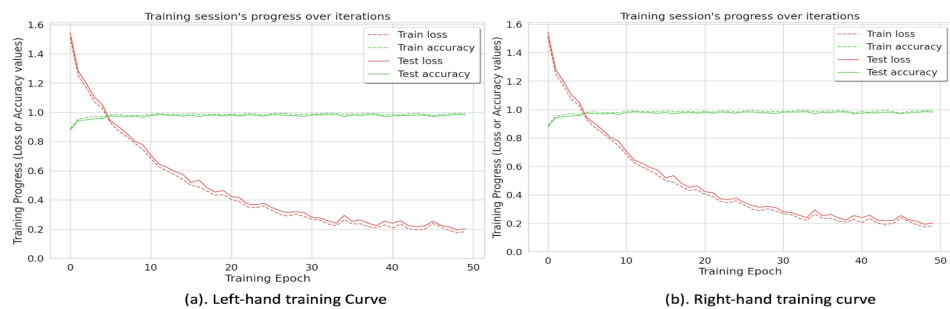

**Figure 3.** The learning curve of the training for left-hand (a) and right-hand (b) (source: authors' experiment results)

## DISCUSSION

The application development results have made it possible to create devices functioning in smartphone and smartwatch environments. Complete data collection is done through a smartwatch. In general, there were no significant difficulties in the data collection process. All data can be sent to the server and re-accessed for use in machine learning processes with LSTM.

The feature selection process shows that there are differences in attributes that affect the right hand and left hand. Considering that all the data were obtained from right-handed students, the feature selection results show that there are several important attributes needed to initiate motions by left-handed people. In general, this indicates that the left hand requires a certain force to initiate a particular motion. This is demonstrated by several attributes indicating rotation detection, such as the Euler sensor, quaternion, and inverse quaternion, all of which are supported by linear accelerating power.

Another interesting fact to discuss is the results on the selected attributes that show that the attributes on the x and z axes are very dominant. The sensors on the y-axis are not selected. These results indicate that the horizontal (x-axis) and the depth (z-axis) motions corresponding to the directions on the device screen - as shown in Figure 4 - are very dominant in the series of exercises performed.

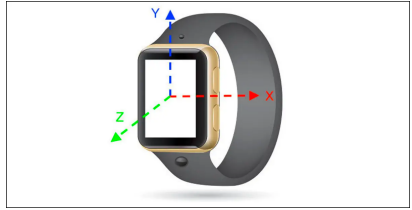

**Figure 4.** Axis direction on the smartwatch screen (source: (Mopas and Huybregts, 2020))

The results of LSTM training show no significant difference in accuracy performance for motions made by the right hand and left hand. This indicates that the motion patterns detected by the sensor correspond to the instructions. The LSTM algorithm also shows very good performance but needs to be validated with more test data. This is one of the next important agendas in this research.

## CONCLUSION

This article describes the development of a smartwatch application to capture indoor exercise motions based on sensors. The software has been used successfully to collect data with specific motions that are useful for muscle strength training. Important attributes that play an important role in the right and left hands are identified as the basis for developing machine learning models with the LSTM algorithm. An important issue for future work is to validate the machine learning models with more data. In addition, it is also necessary to integrate those models into smartphones or smartwatch devices to detect motions automatically. It would be interesting to design a special online community that can encourage each other in physical exercises that are beneficial for maintaining health.

## ACKNOWLEDGMENT

The authors are grateful for the full financial support for this study provided by the Research and Community Service Institute of Universitas Kristen Maranatha (053/SK/ADD/UKM/XI/2022), Indonesia, and Ergo Lab of Industrial Design Department, National Cheng Kung University, Taiwan.

## REFERENCES

Berenguer, A. *et al.* (2016) 'Are smartphones ubiquitous?: An in-depth survey of smartphone adoption by seniors', *IEEE Consumer Electronics Magazine*, 6(1), pp. 104–110.

Birenboim, A. and Shoval, N. (2018) 'Mobility research in the age of the smartphone', *Geographies of Mobility*, pp. 41–49.

Bone, J. *et al.* (2023) 'Leisure engagement in older age is related to objective and subjective experiences of aging'.

Burton, E. *et al.* (2018) 'Reliability and validity of two fitness tracker devices in the laboratory and home environment for older community-dwelling people', *BMC geriatrics*, 18(1), pp. 1–12.

Cooper, C. *et al.* (2018) 'The impact of wearable motion sensing technology on physical activity in older adults', *Experimental gerontology*, 112, pp. 9–19.

*Four Types of Exercise Can Improve Your Health and Physical Ability* (2021) *National Institute on Aging*. Available at: https://www.nia.nih.gov/health/four-types-exercise-can-improve-your-health-and-physical-ability (Accessed: 1 August 2023).

Liu, X.S. (2019) 'A probabilistic explanation of Pearson's correlation', *Teaching Statistics*, 41(3), pp. 115–117.

Maestroni, L. *et al.* (2020) 'The benefits of strength training on musculoskeletal system health: practical applications for interdisciplinary care', *Sports Medicine*, 50(8), pp. 1431–1450.

Meegahapola, L. and Gatica-Perez, D. (2020) 'Smartphone sensing for the well-being of young adults: A review', *IEEE Access*, 9, pp. 3374–3399.

Mopas, M.S. and Huybregts, E. (2020) 'Training by feel: Wearable fitness-trackers, endurance athletes, and the sensing of data', *The Senses and Society*, 15(1), pp. 25–40.

Mylonas, A. *et al.* (2013) 'Smartphone sensor data as digital evidence', *Computers & Security*, 38, pp. 51–75.

Oulasvirta, A. *et al.* (2012) 'Habits make smartphone use more pervasive', *Personal and Ubiquitous computing*, 16, pp. 105–114.

Ramanujam, E., Perumal, T. and Padmavathi, S. (2021) 'Human activity recognition with smartphone and wearable sensors using deep learning techniques: A review', *IEEE Sensors Journal*, 21(12), pp. 13029–13040.

Ronao, C.A. and Cho, S.-B. (2016) 'Human activity recognition with smartphone sensors using deep learning neural networks', *Expert systems with applications*, 59, pp. 235–244.

Schutzer, K.A. and Graves, B.S. (2004) 'Barriers and motivations to exercise in older adults', *Preventive Medicine*, 39(5), pp. 1056–1061. Available at: https://doi.org/10.1016/j.ypmed.2004.04.003.

Steinert, A. *et al.* (2018) 'A wearable-enhanced fitness program for older adults, combining fitness trackers and gamification elements: the pilot study fMOOC@ Home', *Sport Sciences for Health*, 14, pp. 275–282.

Tsapeli, F. and Musolesi, M. (2015) 'Investigating causality in human behavior from smartphone sensor data: a quasi-experimental approach', *EPJ Data Science*, 4(1), p. 24.

*Upper Body Exercises for Seniors and the Elderly, Strength training for seniors,* (2019). Available at: https://www.youtube.com/watch?v=PBMi4Gr_9ls (Accessed: 12 June 2023).

Vooris, R., Blaszka, M. and Purrington, S. (2019) 'Understanding the wearable fitness tracker revolution', *International Journal of the Sociology of Leisure*, 2, pp. 421–437.

Zhu, W. *et al.* (2016) 'Co-occurrence feature learning for skeleton based action recognition using regularized deep LSTM networks', in *Proceedings of the AAAI conference on artificial intelligence*. Available at: https://ojs.aaai.org/index.php/AAAI/article/view/10451 (Accessed: 12 October 2023).